\documentclass{kapproc} % Computer Modern font calls

\usepackage{procps} 

\usepackage[dvips]{graphicx,color}

\upperandlowercase

\normallatexbib %will produce bibliography entries as shown in the
\begin{document}

\articletitle[Bled -- Dipoles]
{Dipole Interactions In Nanosystems}

\author{Philip B. Allen}

%% affil, email, and abstract are optional
\affil{Department of Applied Physics and Applied Mathematics\\
Columbia University, New York, NY 10032\\
and\\
Department of Physics and Astronomy\\
State University of New York, Stony Brook, NY 11794-3800
\footnote{permanent address}}
\email{philip.allen@stonybrook.edu}

%m optional, to supply a shorter version of the title for the running head:
%%\chaptitlerunninghead{}

\anxx{Allen\, Philip B.}

\begin{abstract}
The dipole-dipole interaction influences nanoscopic matter by
fixing the patterns of permanent, displacive, and induced
dipole moments, subject to constraints of molecular
size and other short range interactions.
Prediction of these arrangements is a challenging problem.
The eigenvector of maximum eigenvalue of the dipole-dipole
interaction matrix can provide insights and sometimes
a complete solution.  As an example, the octahedral tilt
instabilities of perovskite-type crystals is shown to 
optimize dipolar interactions.  Therefore this instability
can be designated as antiferroelectric, because dipole-dipole
interactions are a dominant driving force.
\end{abstract}


\section{Introduction}

In condensed matter one finds three kinds of electrical dipoles:
(1) permanent (as in a molecular cluster of polar molecules,
such as CH$_3$Cl), (2) diplacive (as in BaTiO$_3$),
and (3) induced (a cluster of benzene molecules in an
external field).  The unifying feature 
is that the dipole moments feel each other.
Their magnitudes and orientations depend on where the
other dipoles are, in a self-consistent fashion.  This
paper argues that one should study the linear algebra,
and particularly the eigenvector of maximum eigenvalue, of the dipole-dipole
matrix ${\sf \Gamma}$ which relates the induced electric
field $\vec{F}_{i,{\rm ind}}$ at the $i$'th site $\vec{R}_i$
to the dipoles $\vec{\mu}_j$ at sites  $\vec{R}_j$:
\begin{equation}
F_{i\alpha,{\rm ind}}=\sum_{j\beta}\Gamma_{i\alpha,j\beta} \mu_{j\beta},
\label{eq:gammadef}
\end{equation}
\begin{equation}
\Gamma_{i\alpha,j\beta}=\frac{3R_{ij\alpha}R_{ij\beta}-
\delta_{\alpha\beta}R_{ij}^2 } {R_{ij}^5},
\label{eq:gamma}
\end{equation}
where $\vec{R}_{ij}=\vec{R}_i-\vec{R}_j$.
The insights obtained from the eigenvectors and eigenvalues
are illustrated below in some simple examples, and in an
application to  real materials with spontaneous electrical
polarity, the octahedral tilt instability of ReO$_3$-type
perovskite crystals.  

Of course, it must be acknowledged that
not all properties of polar matter have much to do with the
dipole moments of their molecules.  For example, the
water molecule has a dipole in vapor phase, and electrical
polarity is a dominant effect in the 
interactions of water molecules with each other and with
dissolved species.  However, for near-neighbor pairs,
it is not a good approximation to replace the polar charge
distribution of water by a single dipole.  A better 
approximation would use two dipoles, located along
the two bonds, each pointing away from the oxygen toward
one of the two hydrogens.  The sum of these two dipoles
is the total measured dipole moment of the water molecule.
Also, it would be not a very good approximation to assume
that the two dipole moments of the water molecule are
unchanged in their various chemical environments.
Induced polarity should be 
treated on top of permanent or displacive polarity, and
this is done in the case of the octahedral tilt.
A complete theory would not
make any type of dipolar approximation, but would deal
with the actual varying charge distribution of the molecules
in detail.  Modern density functional theory does this
well, but slowly, which motivates a search for
a less complete but still sensible theory.

\section{One Polarizable Molecule}

The energy of a single
molecule is $\mu^2/2\alpha -\vec{\mu} \cdot \vec{F}_{\rm ext}$
where $\alpha$ is the polarizability and
$\vec{F}_{\rm ext}$ is an externally applied field.  The
actual dipole moment $\vec{\mu}$ is fixed by the condition that the
energy is minimum, which gives $\vec{\mu}=\alpha\vec{F}_{\rm ext}$. 

\section{More than One Polarizable Molecule}

Now consider $N$ molecules, 
approximated as points, at chosen fixed positions $\vec{R}_i$.
It costs energy $\mu_i^2/2\alpha$ to create a dipole $\vec{\mu}_i$
on the $i$th site, but one gets back energy $-\vec{\mu}_i \cdot \vec{F}_i$
from the total field $\vec{F}_i=\vec{F}_{i,{\rm ind}}+\vec{F}_{i,{\rm ext}}$
at the site, where $\vec{F}_{i,{\rm ext}}$ 
may be spatially inhomogeneous, varying from site to site.
The total energy, in a standard vector space notation, is
\begin{equation}
U(\vec{\mu}_1,\ldots,\vec{\mu}_N)=
\frac{1}{2}\left<\mu\left|\left(\frac{{\bf 1}}{\alpha}
-{\sf \Gamma}\right)\right|\mu\right> -<\mu|F_{\rm ext}>.
\label{eq:tot}
\end{equation}
When there are $N$ sites, each with a dipole, then $|\mu>$
is a $3N$ column vector containing all the components $\mu_{i\alpha}$
for $\alpha=x,y,z$ and $i=1,\ldots,N$. 
Notice that the interaction energy $-\vec{\mu}_i \cdot \vec{F}_{i,{\rm ind}}$
is reduced by a factor of 2 in the total energy, to avoid double counting.
With no external field, the system is stable if $1/\alpha-{\sf \Gamma}$ is
a positive operator.  Equivalently, all eigenvalues of ${\sf \Gamma}$ should
be less than $1/\alpha$.  

Just as for the single molecule, the induced dipoles $\vec{\mu}_i$
should minimize the total energy, $\partial U/\partial \mu_{i\alpha}=0.$
The solution is
\begin{equation}
|\mu>=\left(\frac{{\bf 1}}{\alpha}
-{\sf \Gamma}\right)^{-1}|F_{\rm ext}>
\label{eq:muind}
\end{equation}
This is a generalized version of the Clausius-Mossotti law \cite{Clausius,
Mossotti,Aspnes}.  An alternate notation,
$|\mu>=\sum (1/\alpha -\gamma)^{-1}<\gamma|F_{\rm ext}> |\gamma>$
uses an expansion in eigenvectors,
where $|\gamma>$ is the eigenvector of ${\sf \Gamma}$ with eigenvalue
$\gamma$.  If the overlap $<\gamma_{\rm max}|F_{\rm ext}>$ is not zero,
the eigenvector of largest eigenvalue $\gamma_{\rm max}$ 
may dominate the response.

If $\gamma_{\rm max}$ increases
to become equal to $1/\alpha$, then the system becomes unstable.
It will develop a spontaneous
polarization, with a pattern $|\mu>$ proportional to the 
corresponding eigenvector $|\gamma_{\rm max}>$.  Higher order terms not
considered here are needed to determine the magnitude of the spontaneous
moment.

\section{Two Dipoles}

For pedagogical purposes, let us examine the
$6\times 6$ matrix ${\sf \Gamma}$  which 
describes the interaction between two dipoles separated
by distance $|\vec{R}|=a$; to be specific, let
$\vec{R}$ point along the $\hat{z}$ axis.  Even without writing 
the matrix, we can find all six eigenvectors by
simple physical reasoning.  The eigenstates of ${\sf \Gamma}$
are those patterns of dipoles such that the induced field
at each dipole is parallel and proportional to the dipole
itself.  Consider the patterns in Fig. 1.{\bf a-d}.  
The field $\vec{F}_1$ lies along the $z$-axis if $\vec{\mu}_2$ points
along $z$, and lies in the $xy$ plane antiparallel to $\vec{\mu}_2$
if $\vec{\mu}_2$ lies in the $xy$ plane.  Therefore vectors
lying along $\hat{z}$ separate from those lying in the $xy$ plane,
and by cylindrical symmetry, eigenvectors in the $xy$ plane
are doubly degenerate.  The degenerate pair can be chosen
so that one lies along $\hat{x}$ and one along $\hat{y}$.
To save space, we project the problem onto the 2d $xz$ plane.
The resulting $4 \times 4$ matrix is
\begin{equation}
\Gamma=\frac{1}{a^3}\left(\begin{array}{rrrr}
0 & 0 & 2 & 0 \\ 
0 & 0 & 0 & -1 \\
2 & 0 & 0 & 0 \\
0 &-1 & 0 & 0 \end{array}\right)
\label{eq:twomat}
\end{equation}
The eigenvalues are (2,1,-1,-2) in units of $(1/a^3)$, and the corresponding
eigenvectors, shown in Fig.1.{\bf a-d}, are
\begin{equation}
|a>=\left(\begin{array}{r} 1 \\ 0 \\ 1 \\ 0 \end{array}\right); \ \ 
|b>=\left(\begin{array}{r} 0 \\-1 \\ 0 \\ 1 \end{array}\right); \ \ 
|c>=\left(\begin{array}{r} 0 \\ 1 \\ 0 \\ 1 \end{array}\right); \ \ 
|d>=\left(\begin{array}{r}-1 \\ 0 \\ 1 \\ 0 \end{array}\right) .
\label{eq:2vecs}
\end{equation}
These eigenvectors all have different symmetries.  Under
inversion and mirror reflection in the central
$xy$ plane, the vectors have $(\lambda_i,\lambda_m)=(--)$, 
$(+-)$, $(-+)$, and $(++)$ symmetry, respectively.

\begin{figure}[ht]
\centerline{\scalebox{0.5}[0.5]{\includegraphics*{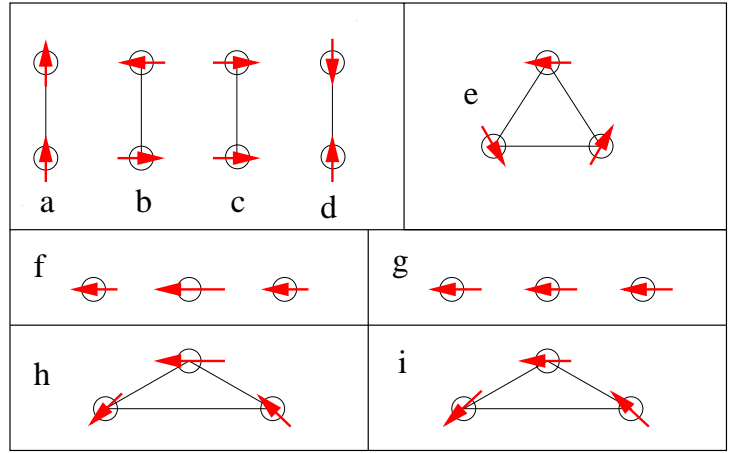}}}
\vskip.2in
\caption{Dipole patterns of two and three interacting dipoles.
In {\bf a-d} the eigenvectors of ${\sf \Gamma}$ are shown for two dipoles.
In {\bf e,f,h}, the eigenvector of largest eigenvalue is shown for
three dipoles.  In {\bf e}, the dipoles all have equal length.  In
{\bf g} and {\bf i} the optimal patterns (not eigenvectors!) for fixed length
dipoles are shown, nearly agreeing with the eigenvectors of 
{\bf f} and {\bf h}.  The vectors in {\bf h} and {\bf i} differ
in angle by about 3$^{\circ}$.}
\end{figure}

The isotropic Heisenberg interaction $U=J\vec{\mu}_1 \cdot \vec{\mu}_2$
for two spins has exactly the same eigenvectors but very different
eigenvalues.  For $J<0$ (ferromagnetic
exchange), states {\bf a}  and {\bf c} are low energy states, equal
in energy, and states {\bf b} and {\bf d} are high energy states,
also equal in energy.  For $J<0$ (antiferromagnetic exchange),
the reverse is true.  For magnetic systems (the O$_2$ molecule
is a nice example), the exchange energy is much larger than
the dipole-dipole interaction.  The ground state of O$_2$
has two outer electrons, polarized ferromagnetically ($S=1$), and only a 
tiny spin-orbit-induced preference for the moment to point
perpendicular to the axis.  By contrast, electrical dipoles
have a large dipole-dipole interaction which is intrinsically
anisotropic and strongly prefers state {\bf a}.

\section{Three Dipoles}

In the two-dipole problem of the previous section,
the eigenstates all had the property that each
molecule's moment $\vec{\mu}_i$ (where $\mu_{i\alpha}=<i\alpha|\gamma>$)
had the same magnitude $|\vec{\mu}_i|$.  This brings a nice
benefit: these patterns tell us also about the energetics of
the problem of fixed size permanent dipoles.

Fig. 1.{\bf e-i} shows geometries and corresponding
dipole patterns for three molecules.  The physical situations of
interest are (1) polarizable molecules with
no permanent moment, (2) permanent dipoles
which can orient in space but not change their magnitude,
and (3) permanent dipoles which are also polarizable and
change their magnitude.  We omit category (3) for now.
Category (1) is mathematically simplest because linear
algebra determines what happens in an external field, and
whether spontaneous polarity appears.  Fig. 1.{\bf e,f,h} show
the eigenvectors of largest eigenvalue of ${\sf \Gamma}$
for ({\bf e}) the equilateral triangle, ({\bf f}) the linear arrangement,
and ({\bf h}) an obtuse isosceles triangle.  For the equilateral
case, the eigenvector is determined by symmetry, but for the 
other two cases, there are two independent vectors odd
under the perpendicular mirror, so a $2 \times 2$ matrix
is needed, and the central atom (or molecule)
has a larger $|\vec{\mu}_{\rm ind}|$ than the end atoms.

Because $|\vec{\mu}_i|$ is no longer the same for
each molecule in the maximal eigenvector, it is
not the solution of the problem of a fixed moment.  Instead,
for fixed moments we want to know
the pattern $|\mu>$ which minimizes
\begin{equation}
U(\vec{\mu}_1,\ldots,\vec{\mu}_N)=
-\frac{1}{2}<\mu|{\sf \Gamma}|\mu> -<\mu|F_{\rm ext}>,
\label{eq:totf}
\end{equation}
subject to $N$ nonlinear constraints $|\vec{\mu}_i|=\mu_0$.
This is a much harder mathematical problem.  Fortunately,
the solution is guaranteed \cite{Allen} to have a smooth relation
to the maximal eigenvector of ${\sf \Gamma}$.  The 
solutions to the fixed moment problem in zero field
are shown in Fig. 1.{\bf g,i}, and are close to the
unconstrained solutions in Fig. 1.{\bf f,h}.

\section{Infinite Stack of Dipoles}

\begin{figure}[h]
\centerline{\scalebox{0.3}[0.3]{\includegraphics*{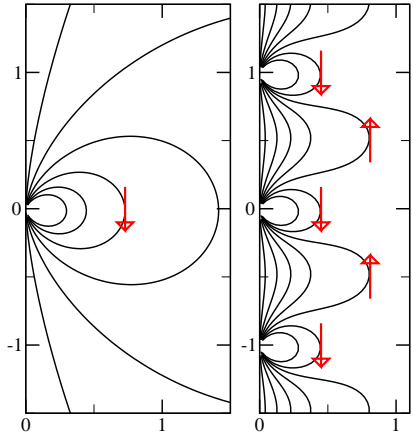}}}
\vskip.2in
\caption{Faraday flux lines indicating the field strength
and direction (indicated by arrows) for a single dipole pointing up (left panel)
and for a periodic stack of dipoles pointing up (right panel).  In
the left panel the field falls as $1/r^3$ in all directions.
In the right, the field decays exponentially in the directions
transverse to the stack.}
\end{figure}

Consider the arrangement $|\mu>$ of Fig. 2, with equally spaced
dipoles $\vec{\mu}$ of equal magnitude, all pointing in the
$\hat{z}$ direction which is also the
direction of the stack of molecules.  It is clear that this
pattern is the maximal eigenvector of the corresponding ${\sf \Gamma}$.
The eigenvalue is $4\zeta(3)/a^3$ where $a$ is the spacing
of dipoles and $\zeta(3)=1.202057\ldots$ is the Riemann zeta function.
Since the moments are equal, this pattern is also the solution
to the problem of fixed dipoles on a 1-d line.  It is interesting
to compute the electric field of this stack, which is the negative
gradient of the electrostatic potential $\Phi(\rho,z)$ (where
$\rho = \sqrt{x^2 +y^2}$ is the radial distance).
By methods described elsewhere
\cite{Allen} the potential has the asymptotic form
\begin{equation} 
\Phi(\rho,z)=-\frac{4\pi\mu}{a^2} \frac{\sin(2\pi z/a)e^{-2\pi \rho/a}}
{\sqrt{\rho/a}}
\label{eq:phi}
\end{equation}
This result explains why fixed dipoles particularly like to form
one-dimensional stacks.   The dipole-dipole attraction within
the stack is very big, and the interaction with other dipoles falls
very rapidly with distance.  The asymptotic form Eq.\ref{eq:phi}
is already accurate to 2\% at $\rho=a$.

\section{Cubic lattice of Dipoles}

\begin{figure}[ht]
\centerline{\scalebox{0.4}[0.4]{\includegraphics*{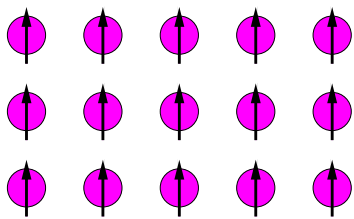}}}
\vskip.2in
\caption{Infinite periodic lattice of cubic symmetry, showing the
ferroelectric eigenstate studied by Clausius and Mossotti.  Also
the 2-d analog is an eigenstate.  These are not necessarily
the maximal eigenstates.}
\end{figure}

An eigenvector of this problem was found independently by
Clausius \cite{Clausius} and Mossotti \cite{Mossotti}, and
is illustrated in Fig. 3.  If all dipoles point along a cube
axis, with equal magnitude, it is clear that the induced
field does the same thing, so this is an eigenvector.  The
Clausius-Mossotti arguments show that the eigenvalue is
$\gamma_{\rm CM}=4\pi n/3$ where $n$, the density of 
molecules, is $1/a^3$ for the simple cubic lattice.
A good pedagogical discussion is given by Aspnes \cite{Aspnes}.
One should also ask whether this is the maximal eigenvalue.
From Fig. 2, it is clear that for the simple cubic lattice
illustrated in Fig. 3, the answer is ``no.''  One can
further lower the energy by reversing the direction of
alternate stacks of dipoles, producing a slightly larger
eigenvalue in an antiferroelectric pattern.  Luttinger
and Tisza \cite{Luttinger} proved that this pattern
had the extremal eigenvalue.  For diamond structure,
a more complicated extremal antiferroelectric eigenstate
was found by White {\it et al.} \cite{White}.  These authors
also showed that for {\it fcc} and {\it bcc} structures,
the Clausius-Mossotti ferroelectric eigenstate is maximal.

\section{Octahedral Tilt}

\begin{figure}[ht]
\centerline{\scalebox{0.35}[0.35]{\includegraphics*{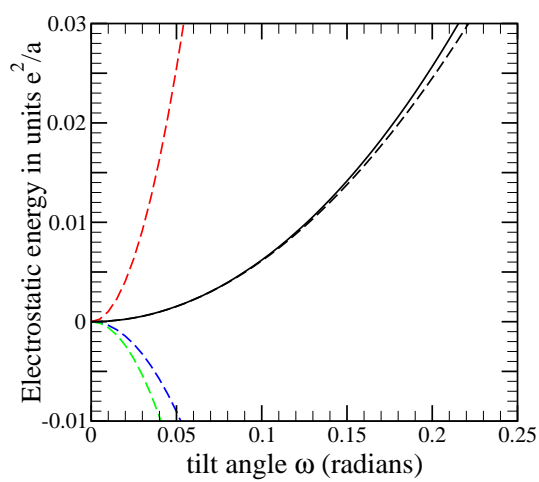}}}
\vskip.2in
\caption{Energy versus tilt angle for $(\pi,\pi,\pi)$ periodic tilts
around the (111) axis of BX$_3$ (ReO$_3$ structure type) perovskites.
The solid line is the total electrostatic energy (units $Z^2 e^2/a$,
where $Z$ is the anion charge) computed for rigid ions comprising
rigid octahedra, with a corresponding hard-core contraction perpendicular to
(111) planes.  The dashed curves , from top to bottom,
are electrostatic energy for one displaced anion, dipole-dipole attraction
energy of interacting displacements, and volume contraction energy.  The
sum of these is the dashed curve, coinciding with the exact energy
up to fourth order terms in the Taylor series.}
\end{figure}

In this last section a brief discussion is given of recent work
to be explained more fully elsewhere \cite{Allen2,Chaudhuri,Chen}.  
We have found that dipole-dipole interactions play an
unexpectedly big role in the ``octahedral tilt'' instabilities
common in perovskite materials \cite{Glazer,Woodward,Howard,Thomas}.
The motif is evident in Fig. 1.{\bf e}, showing the tendency
of a triangular plaquette to polarize circumferentially.  In the
perovskite tilt, when the axis is (111), the displaced anions
create the same triangular pattern of displacive dipoles.  
The moment points from the new
anion position to the undistorted cubic anion position.  These
moments interact because they create an induced field at each
undisplaced site.  After some thought one can conclude that this
is an eigenstate of ${\sf \Gamma}$, and after some calculation one
can prove that it is the maximal eigenstate.  There is no better
way to place dipoles on the X sites of ABX$_3$ perovskite than
in the pattern corresponding to this tilt.  The eigenvalue
$\gamma_{\rm max}=14.461/a^3$, is 15\% bigger than $\gamma_{\rm CM}$.
The consequence is that in a rigid ion picture, the BX$_3$ simplification
of the ABX$_3$ lattice is only marginally stable under electrostatic
and hard sphere forces.  When the anion polarizability is added to
the theory \cite{Allen2}, the structure is almost immediately unstable even for
very small polarizability.  The maximum eigenvalue is accidentally
5-fold degenerate.  Three of the partners permit $(\pi,\pi,\pi)$
alternating tilts in any linear combination around any cube axis.  
These generate the distortions labeled $a^-b^-c^-$ in Glazer
notation \cite{Glazer}.  The Glazer distortions with + signs
require tilts with $(\pi,\pi,0)$ periodicity, which happen also
to be eigenvectors of ${\sf \Gamma}$, with slightly smaller
than maximal eigenvalue.  
It seems appropriate, given the role that dipolar
interactions play, to regard all these distorted ground states
as multiple-sublattice antiferroelectrics.  The
other two partners are related to the cooperative Jahn-Teller  
ground state of LaMnO$_3$, which thus also has an
unexpected contribution from polar interactions in its distorted
ground state.

\section{Summary}

Maximal eigenstates of the dipole-dipole
interaction tensor ${\sf \Gamma}$ are useful and relevant to 
several problems: response of polarizable clusters to external fields,
causes and patterns of spontaneous distortion of polar assemblies,
and the self-organization of polar molecules in various media.
It is surprising how robust and useful this concept
seems to  be.  Given that short range repulsive forces are 
large, and that near-neighbor electrostatics may involve higher multipoles,
there is no obvious reason why a pattern like the perovskite octahedral
tilt should have to conform to the maximal eigenvector.  The fact that
it does should encourage us to use this tool in other problems.

\begin{acknowledgments}
I thank Dr. Y.-R. Chen, C. P. Grey, and S. Chaudhuri for 
help and stimulation.
The work was supported in part by NSF grant no. DMR-0089492, and
in part by a J. S. Guggenheim Foundation fellowship.  Work at
Columbia was supported in part by the MRSEC Program of the NSF
under award no. DMR-0213574.

\end{acknowledgments}

\end{document}